\documentclass[aps,prl,twocolumn,superscriptaddress]{revtex4-1}
\usepackage{graphicx,latexsym}
\usepackage{amsmath,amssymb,amsfonts}
\usepackage{bm}
\usepackage{braket}
\newcommand{\up}{\uparrow}
\newcommand{\down}{\downarrow}

\begin{document}

%\preprint{}

\title{Dynamical Sign Reversal of Magnetic Correlations in Dissipative Hubbard Models}
\author{Masaya Nakagawa}
\email{nakagawa@cat.phys.s.u-tokyo.ac.jp}
\affiliation{Department of Physics, University of Tokyo, 7-3-1 Hongo, Bunkyo-ku, Tokyo 113-0033, Japan}
\author{Naoto Tsuji}
\affiliation{RIKEN Center for Emergent Matter Science (CEMS), Wako, Saitama 351-0198, Japan}
\author{Norio Kawakami}
\affiliation{Department of Physics, Kyoto University, Kyoto 606-8502, Japan}
\author{Masahito Ueda}
\affiliation{Department of Physics, University of Tokyo, 7-3-1 Hongo, Bunkyo-ku, Tokyo 113-0033, Japan}
\affiliation{RIKEN Center for Emergent Matter Science (CEMS), Wako, Saitama 351-0198, Japan}
\affiliation{Institute for Physics of Intelligence, University of Tokyo, 7-3-1 Hongo, Bunkyo-ku, Tokyo 113-0033, Japan}


%\author{}
%\thanks{}
%\affiliation{}


\date{\today}

\begin{abstract}
In quantum magnetism, the virtual exchange of particles mediates an interaction between spins. 
Here, we show that an inelastic Hubbard interaction fundamentally alters the magnetism of the Hubbard model due to dissipation in spin-exchange processes, leading to sign reversal of magnetic correlations in dissipative quantum dynamics. 
This mechanism is applicable to both fermionic and bosonic Mott insulators, and can naturally be realized with ultracold atoms undergoing two-body inelastic collisions. 
The dynamical reversal of magnetic correlations can be detected by using a double-well optical lattice or quantum-gas microscopy, the latter of which facilitates the detection of the magnetic correlations in one-dimensional systems because of spin-charge separation. 
Our results open a new avenue toward controlling quantum magnetism by dissipation.  
\end{abstract}

%\keywords{}

\maketitle

%%%%%[Introduction]%%%%%

Quantum magnetism in Mott insulators is one of the central problems in strongly correlated many-body systems \cite{Auerbach}. A Mott insulator is described by the Hubbard model, where a strong repulsive interaction between particles precludes multiple occupation and anchors a single spin to each lattice site. 
While the kinetic motion of particles is frozen in Mott insulators, quantum mechanics allows particles to virtually hop between sites. A second-order process involving virtual exchange of particles leads to an effective spin-spin interaction, providing the fundamental origin of quantum magnetism \cite{Auerbach}. Recent developments in quantum simulations of the Hubbard model with ultracold atoms \cite{Esslinger10} have offered a powerful approach to unveiling low-temperature properties of quantum magnets \cite{Trotzky08, Greif13, Greif15, Hart15, Ozawa18}. In particular, quantum-gas microscopy has enabled site-resolved imaging of spin states \cite{Bakr09, Parsons15, Cheuk15, Cheuk16_1}, culminating in direct observation of antiferromagnetic correlations and long-range order in the Hubbard model \cite{Parsons16, Cheuk16_2, Boll16, Mazurenko17}. 
The essential requirement for observing the quantum magnetism is to achieve sufficiently low temperatures comparable with the exchange coupling. 

In this Letter, we demonstrate that ultracold atoms undergoing inelastic collisions obey a completely different principle for realizing quantum magnetism; instead of relaxing to low-energy states, those atoms stabilize \textit{high-energy} states due to dissipation caused by inelastic collisions. 
Inelastic collisions have widely been observed for atoms in excited states \cite{Sponselee18, Tomita18} and molecules \cite{Syassen08, Zhu14}, and can be artificially induced by photoassociation \cite{Tomita17}. 
In contrast to standard equilibrium systems that favor low-energy states, the long-time behavior of dissipative systems is governed by the lifetime of each state under dissipation. 
We show that the spin-exchange mechanism is altered in the presence of inelastic collisions due to dissipation in an intermediate state. As a result, dissipation dramatically changes the magnetism of the Hubbard model; the magnetism is \textit{inverted} from the conventional equilibrium one, leading to the sign reversal of spin correlations through dissipative dynamics.

%%%%%%%%%%%%%%%%%%%%%%%%%
\begin{figure}
\includegraphics[width=8.5cm]{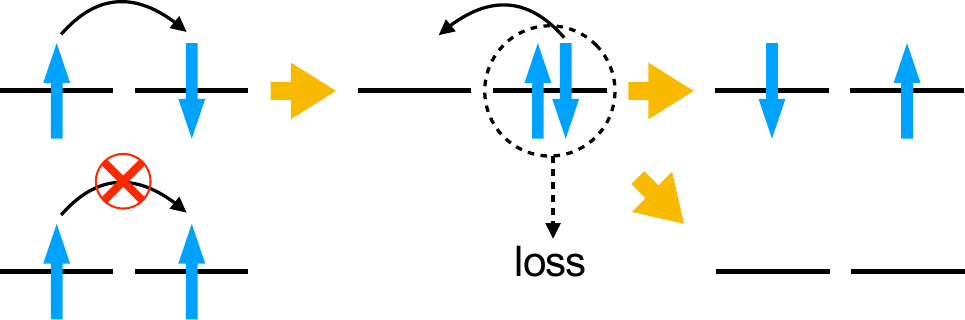}
\caption{Schematic illustration of a second-order process mediating the spin-exchange interaction in the dissipative Fermi-Hubbard system. A loss in an intermediate process causes a finite lifetime of the system.}
\label{fig_exchange}
\end{figure}
%%%%%%%%%%%%%%%%%%%%%%%%%

The spin-exchange interaction in the presence of an inelastic interaction, which plays a key role in this Letter, is schematically illustrated in Fig.~\ref{fig_exchange} for the Fermi-Hubbard system. Since an intermediate state in the second-order process involves a doubly occupied site, an antiferromagnetic spin configuration has a finite lifetime due to a particle loss in the intermediate state, whereas a ferromagnetic spin configuration cannot decay due to the Pauli exclusion principle. Because of this dissipative spin-exchange interaction, low-energy states gradually decay, and high-energy spin states will eventually be stabilized. 
Such stabilization of high-energy states cannot be achieved in conventional equilibrium systems and is reminiscent of negative-temperature states \cite{LandauLifshitz, Ramsey56} realized in isolated systems \cite{Purcell51, Hakonen92, Hakonen94, Rapp10, Rapp12, Tsuji11, Braun13, Gauthier18, Johnstone18, Yamamoto19}. 
In contrast, here dissipation to an environment plays a vital role and thus offers a unique avenue towards the control of magnetism in open systems.

%%%%%[Setup]%%%%%

\textit{Model}.--\ 
We consider a dissipative Hubbard model of two-component fermions or bosons realized with ultracold atoms in an optical lattice. The unitary part of the dynamics is governed by the Hubbard Hamiltonian which reads
\begin{equation}
H=-t\sum_{\langle i,j\rangle,\sigma=\up,\down}(c_{i\sigma}^\dag c_{j\sigma}+\mathrm{H.c.})+U\sum_j n_{j\up}^{(f)}n_{j\down}^{(f)}
\label{eq_FH}
\end{equation}
for fermions, and
\begin{align}
H=&-t\sum_{\langle i,j\rangle,\sigma=\up,\down}(b_{i\sigma}^\dag b_{j\sigma}+\mathrm{H.c.})+\sum_j U_{\up\down}n_{j\up}^{(b)}n_{j\down}^{(b)}\notag\\
&+\sum_\sigma\sum_j \frac{U_{\sigma\sigma}}{2}n_{j\sigma}^{(b)}(n_{j\sigma}^{(b)}-1)
\label{eq_BH}
\end{align}
for bosons. Here $c_{j\sigma}$ ($b_{j\sigma}$) is the annihilation operator of a fermion (boson) with spin $\sigma$ at site $j$, and $n_{j\sigma}^{(f)}=c_{j\sigma}^\dag c_{j\sigma}$ ($n_{j\sigma}^{(b)}=b_{j\sigma}^\dag b_{j\sigma}$). 
We assume that hopping with an amplitude $t$ occurs between the nearest-neighbor sites and that the on-site elastic interactions are repulsive: $U,U_{\sigma\sigma'}>0$. We also assume $t>0$ without loss of generality.  
Now we suppose that atoms also undergo inelastic collisions; because a large internal energy is converted to the kinetic energy, two atoms after inelastic collisions quickly escape from the trap and are lost. 
The dissipative dynamics of the density matrix $\rho$ of the system at time $\tau$ is described by the following quantum master equation \cite{BreuerPetruccione}:
\begin{equation}
\frac{d\rho}{d\tau}=i[\rho,H]+\sum_{j,\sigma,\sigma'}\left(L_{j\sigma\sigma'}\rho L_{j\sigma\sigma'}^\dag-\frac{1}{2}\{ L_{j\sigma\sigma'}^\dag L_{j\sigma\sigma'},\rho\}\right).
\label{eq_master}
\end{equation}
The Lindblad operators $L_{j\sigma\sigma'}$ induce two-body losses due to the on-site inelastic collisions, and are expressed as $L_{j\sigma\sigma'}=\sqrt{2\gamma}c_{j\sigma}c_{j\sigma'}\delta_{\sigma,\uparrow}\delta_{\sigma',\downarrow}$ for fermions and $L_{j\sigma\sigma'}= \sqrt{\gamma_{\sigma\sigma'}}b_{j\sigma}b_{j\sigma'}$ for bosons. The coefficients $\gamma,\gamma_{\sigma\sigma'}>0$ are determined from the loss rates of atoms.

%%%%%[Basic mechanism]%%%%%

\textit{Spin-exchange interaction in dissipative systems}.--\ 
We first illustrate the basic mechanism that underlies the magnetism of the dissipative Hubbard systems. 
We consider a strongly correlated regime ($U,U_{\sigma\sigma'}\gg t$) and assume that the initial particle density is set to unity so that a Mott insulating state is realized. 
For simplicity, we consider the case of the spin SU(2) invariance, i.e., $U_{\up\up}=U_{\down\down}=U_{\up\down}=U$. 
Then, if doubly occupied states and empty states are ignored, the Fermi (Bose) Hubbard model \eqref{eq_FH} [\eqref{eq_BH}] reduces to the antiferromagnetic (ferromagnetic) Heisenberg model $H_{\mathrm{spin}}=J\sum_{\langle i,j\rangle}(\bm{S}_i\cdot\bm{S}_j-1/4)$ [$H_{\mathrm{spin}}=-J\sum_{\langle i,j\rangle}(\bm{S}_i\cdot\bm{S}_j+3/4)$] with the spin-exchange interaction $J=4t^2/U$ \cite{Kuklov03, Duan03}. 

Here we employ the quantum-trajectory method \cite{Dalibard92, Carmichael_book, Daley14} to investigate the dynamics described by Eq.\ \eqref{eq_master} \cite{supple}. The dynamics is decomposed into a nonunitary Schr\"{o}dinger evolution under an effective non-Hermitian Hamiltonian $H_{\mathrm{eff}}\equiv H-\frac{i}{2}\sum_{j,\sigma,\sigma'} L_{j\sigma\sigma'}^\dag L_{j\sigma\sigma'}$ and stochastic quantum-jump processes which induce particle losses with the jump operators $L_{j\sigma\sigma}$. The non-Hermitian Hamiltonian $H_{\mathrm{eff}}$ is obtained if we replace the Hubbard interactions $U$ and $U_{\sigma\sigma'}$ with $U-i\gamma$ and $U_{\sigma\sigma'}-i\gamma_{\sigma\sigma'}$, respectively, thereby making the interaction coefficients complex-valued due to the inelastic interactions. In each quantum trajectory, the system evolves under the non-Hermitian Hubbard model during a time interval between loss events \cite{Ashida16, Ashida18}. 
Each quantum trajectory is characterized by the number of loss events. 
Let us first consider a trajectory that does not involve any loss event; along this trajectory, the particle number stays constant. 
Since the double occupancy is still suppressed due to the large Hubbard interaction $U$, the dynamics is constrained to the Hilbert subspace of the spin Hamiltonian. The effective spin Hamiltonian, which governs the dynamics in the quantum trajectory, is derived from the non-Hermitian Hubbard model through the second-order perturbation theory, giving
\begin{equation}
H_{\mathrm{eff}}=\eta(J_{\mathrm{eff}}+i\Gamma)\sum_{\langle i,j\rangle}\left(\bm{S}_i\cdot\bm{S}_j+\frac{1-2\eta}{4}\right),
\label{eq_effHeis}
\end{equation}
where $J_{\mathrm{eff}}=4Ut^2/(U^2+\gamma^2)$, $\Gamma=4\gamma t^2/(U^2+\gamma^2)$, and $\eta=+1$ ($\eta=-1$) for fermions (bosons).  Here we assume spin-independent dissipation $\gamma_{\sigma\sigma'}=\gamma$ for bosonic atoms (see Supplemental Material \cite{supple} for a general case). 
Equation~\eqref{eq_effHeis} shows that the spin-spin interactions are affected by dissipation even if the double occupancy is suppressed by the strong repulsion, 
since the virtual second-order process involves a doubly occupied site (see Fig.~\ref{fig_exchange}). In fact, the energy denominators in $J_{\mathrm{eff}}=\mathrm{Re}[4t^2/(U-i\gamma)]$ and $\Gamma=\mathrm{Im}[4t^2/(U-i\gamma)]$ reflect the dissipation in the intermediate state. The eigenenergy of the Hamiltonian \eqref{eq_effHeis} is given by $E_n=(J_{\mathrm{eff}}+i\Gamma)E_n^{(0)}/J$, where $E_n^{(0)}\leq 0$ is the eigenenergy of the Heisenberg Hamiltonian $H_{\mathrm{spin}}$. Thus, the decay rate of the $n$th eigenstate, which is given by the imaginary part of the energy, is proportional to $E_n^{(0)}$: $-\mathrm{Im}[E_n]=-(\Gamma/J)E_n^{(0)}\geq 0$. Since $E_n^{(0)}\leq 0$, this indicates that lower-energy states have larger decay rates with shorter lifetimes. Therefore, after a sufficiently long time, only the high-energy spin states survive. 
This implies that the dissipative Fermi (Bose) Hubbard system develops ferromagnetic (antiferromagnetic) correlations. The mechanism can intuitively be understood from Fig.~\ref{fig_exchange} for the Fermi-Hubbard system as no decay occurs for a ferromagnetic spin configuration. For the Bose-Hubbard system, while additional spin-exchange processes due to the absence of the Pauli exclusion principle for ferromagnetic spin configurations lead to a ferromagnetic Heisenberg interaction for a closed system, dissipation during the exchange processes renders ferromagnetic states to decay faster than antiferromagnetic states in the dissipative system.

%%%%%[Double-well systems]%%%%%

%%%%%%%%%%%%%%%%%%%%%%%%%
\begin{figure}
\includegraphics[width=8.5cm]{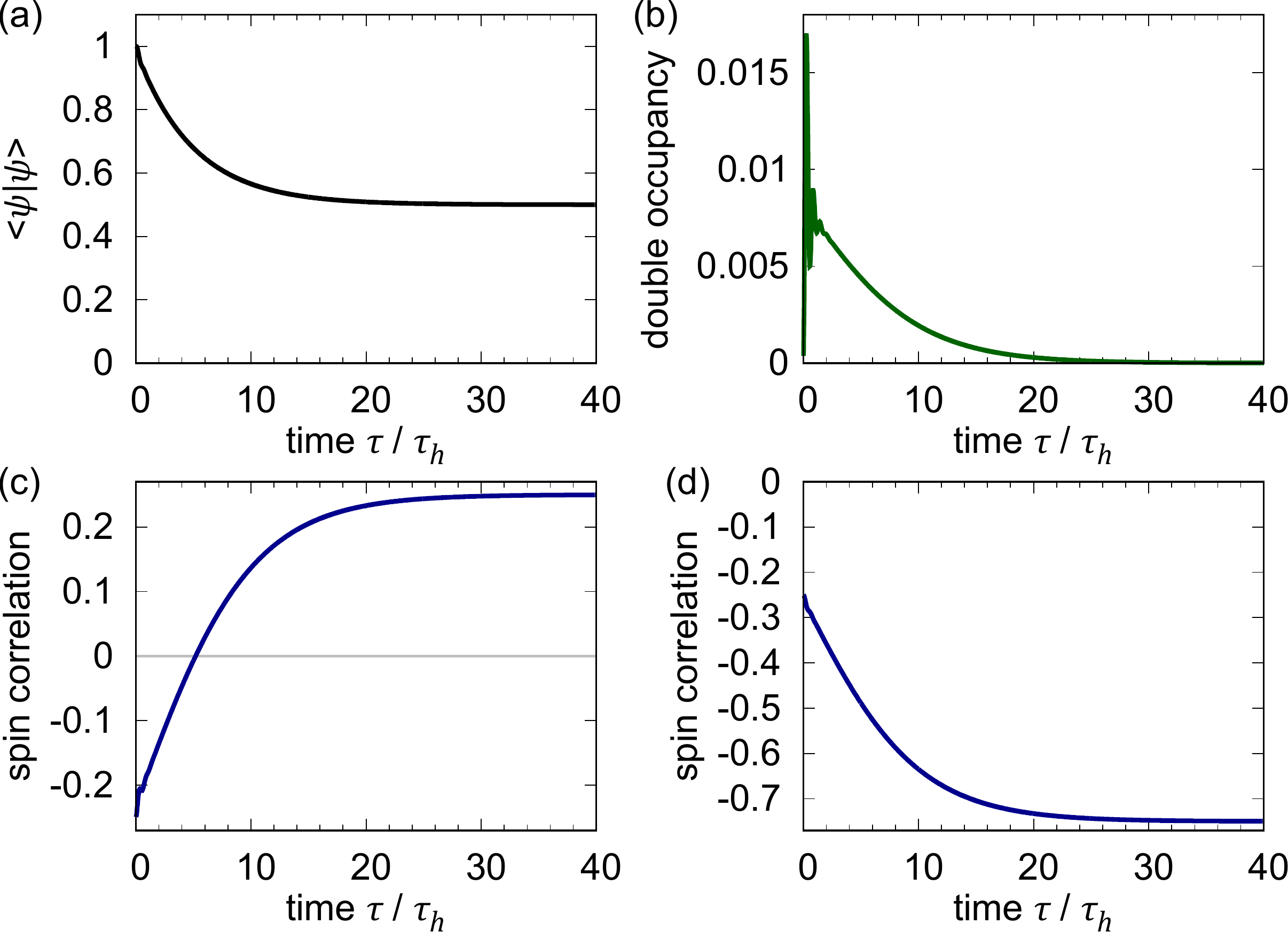}
\caption{(a) Time evolution of the squared norm $\braket{\psi(\tau)|\psi(\tau)}$ 
and (b) that of the double occupancy $\bra{\psi(\tau)}\frac{1}{2}(n_{1\up}^{(a)}n_{1\down}^{(a)}+n_{2\up}^{(a)}n_{2\down}^{(a)})\ket{\psi(\tau)}/\braket{\psi(\tau)|\psi(\tau)}$ ($a=f$ or $b$). Note that these quantities take the same values for the Fermi and Bose Hubbard models. 
(c) Time evolution of the spin correlation $\bra{\psi(\tau)}\bm{S}_1\cdot\bm{S}_2\ket{\psi(\tau)}/\braket{\psi(\tau)|\psi(\tau)}$ 
of the Fermi-Hubbard model and (d) that of the Bose-Hubbard model. The parameters are set to $U/t=10$ and $\gamma/t=3$. The unit of time is the inverse hopping rate $\tau_h=1/t$.}
\label{fig_2site}
\end{figure}
%%%%%%%%%%%%%%%%%%%%%%%%%

\textit{Double-well systems}.--\ 
A minimal setup to demonstrate the basic principle described above is a two-site system. It can be experimentally realized with an ensemble of double wells created by optical superlattices \cite{Trotzky08, Greif13}, and magnetic correlations between the left and right wells can be measured from singlet-triplet oscillations \cite{Greif13, Greif15, Ozawa18}. We consider an ensemble of double wells in which two particles with opposite spins occupy each double well. 
During the dissipative dynamics, a double well in which a loss event takes place becomes empty. Therefore, when a magnetic correlation is measured at time $\tau$, signals come from those double wells in which particles have not yet been lost. Such wells are faithfully described by the quantum trajectory without loss events. 

Figure~\ref{fig_2site} shows the time evolutions of (a) the squared norm of the state $\braket{\psi(\tau)|\psi(\tau)}$, (b) the double occupancy $\bra{\psi(\tau)}\frac{1}{2}(n_{1\up}^{(a)}n_{1\down}^{(a)}+n_{2\up}^{(a)}n_{2\down}^{(a)})\ket{\psi(\tau)}/\braket{\psi(\tau)|\psi(\tau)}\ (a=f,b)$, and (c) (d) the spin correlation $\bra{\psi(\tau)}\bm{S}_1\cdot\bm{S}_2\ket{\psi(\tau)}/\braket{\psi(\tau)|\psi(\tau)}$ obtained from a numerical solution of the Schr\"{o}dinger equation $i\partial_\tau\ket{\psi(\tau)}=H_{\mathrm{eff}}\ket{\psi(\tau)}$. Here $H_{\mathrm{eff}}$ is the two-site non-Hermitian Fermi (Bose) Hubbard model and the initial state is assumed to be $c_{1\up}^\dag c_{2\down}^\dag\ket{0}$ ($b_{1\up}^\dag b_{2\down}^\dag\ket{0}$), where $\ket{0}$ is the particle vacuum. The results clearly show that the dissipative Fermi (Bose) Hubbard system develops a ferromagnetic (antiferromagnetic) correlation which is eventually saturated at $0.25$ ($-0.75$), indicating a formation of 
the highest-energy spin state $(\ket{\up}_1\ket{\down}_2+\ket{\down}_1\ket{\up}_2)/\sqrt{2}$ [$(\ket{\up}_1\ket{\down}_2-\ket{\down}_1\ket{\up}_2)/\sqrt{2}$] of the Heisenberg model. We note that the double occupancy in the dynamics is almost negligible and further suppressed by an increase in the dissipation $\gamma$ 
(see Supplemental Material for the dependence on $\gamma$ \cite{supple}); 
the latter is due to the continuous quantum Zeno effect \cite{Syassen08, Zhu14, Tomita17} by which strong dissipation inhibits hopping to an occupied site. Nevertheless, virtual hopping is allowed, leading to the growth in the spin correlation. 

Another important feature is that the squared norm stays constant after the spin correlation is saturated. Since the squared norm corresponds to the probability of the lossless quantum trajectory \cite{Daley14}, the saturation signals that the system enters a dark state that is immune to dissipation. This property explains why the highest-energy spin state is realized in the long-time limit; the spin-symmetric (spin-antisymmetric) state of fermions (bosons) is actually free from dissipation and thus has the longest lifetime, since in this spin configuration both Fermi and Bose statistics dictate antisymmetry of the real-space wave function and thus allow no double occupancy \cite{FossFeig12}.

%%%%%[Bulk systems]%%%%%

\textit{Extracting spin correlations from conditional correlators}.--\ 
Having established the basic mechanism of the magnetism induced by dissipation, 
we now include the effect of quantum jumps, which create holes due to particle loss. 
One might think that the created holes scramble the background spin configuration and disturb the development of the spin correlation. 
Below we show that this issue can be circumvented by using quantum-gas microscopy for the one-dimensional Hubbard models. 

\begin{figure}
\includegraphics[width=8.5cm]{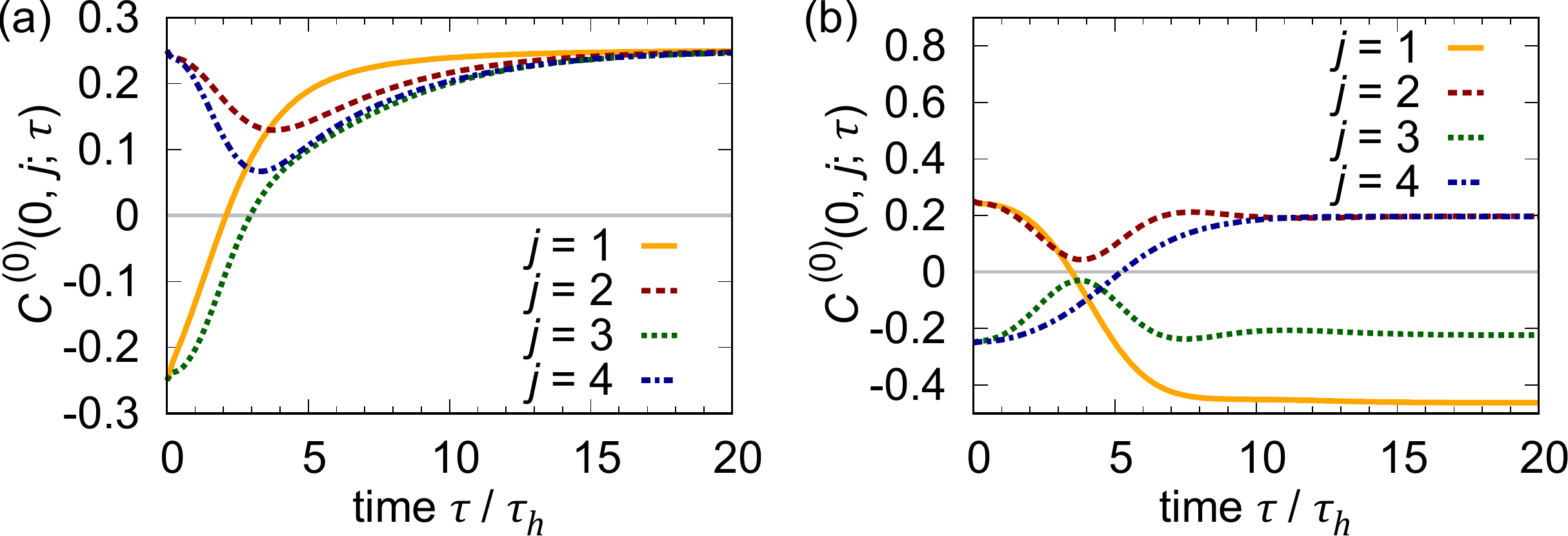}
\caption{Dynamics of the spin correlations $C^{(0)}(0,j;\tau)$ for the dissipative 8-site Fermi [(a)] and 6-site Bose [(b)] Hubbard systems in the absence of quantum-jump events. The parameters are set to $U/t=10$ and $\gamma/t=10$. The unit of time is the inverse hopping rate $\tau_h=1/t$.}
\label{fig_nojump}
\end{figure}

We first show in Fig.~\ref{fig_nojump} the time evolution of the one-dimensional dissipative Hubbard models in quantum trajectories without quantum-jump events. The system size is $N=8$ ($N=6$) for the Fermi (Bose) system. 
The initial states are chosen to be a N\'{e}el state $\ket{\up\down\up\down\up\down\up\down}$ for the Fermi system, and a ferromagnetic domain-wall state $\ket{\up\up\up\down\down\down}$ for the Bose system, in accordance with the equilibrium spin configuration of each system without dissipation. After the dissipation is switched on at $\tau=0$, the Fermi (Bose) system in Fig.~\ref{fig_nojump} (a) [Fig.~\ref{fig_nojump} (b)] clearly develops a ferromagnetic (antiferromagnetic) spin correlation $C^{(0)}(i,j;\tau)\equiv\bra{\psi(\tau)}\bm{S}_i\cdot\bm{S}_j\ket{\psi(\tau)}/\braket{\psi(\tau)|\psi(\tau)}$, whose sign is reversed from that of the initial state, and the correlation is eventually saturated at a value consistent with the highest-energy state of the antiferromagnetic (ferromagneic) Heisenberg chain. 

%%%%%%%%%%%%%%%%%%%%%%%%%
%%%%%%%%%%%%%%%%%%%%%%%%%

Quantum-gas microscopy enables a high-precision measurement of the particle number at the single-site resolution \cite{Bakr09, Parsons15, Cheuk15, Cheuk16_1}. Given a single-shot image of an atomic gas, the occupation number of each site is identified to be zero, one, or two. 
From this information, one can find the number of quantum jumps that have occurred by the time of the measurement. 
Accordingly, one can take an ensemble average over quantum trajectories with a given number of quantum jumps \cite{Ashida18_2}. 
The density matrix conditioned on the number of quantum jumps from the initial time to $\tau$ is given by $\rho^{(n)}(\tau)\equiv\mathcal{P}^{(n)}\rho(\tau)\mathcal{P}^{(n)}/\mathrm{Tr}[\mathcal{P}^{(n)}\rho(\tau)\mathcal{P}^{(n)}]$. Here $\mathcal{P}^{(n)}$ is a projector onto the sector in which $n$ quantum jumps have occurred. Then, one can calculate the correlation function $C^{(n)}(i,j;\tau)\equiv \mathrm{Tr}[\rho^{(n)}(\tau)\bm{S}_i\cdot\bm{S}_j]$ \cite{supple}.

\begin{figure}
\includegraphics[width=8.5cm]{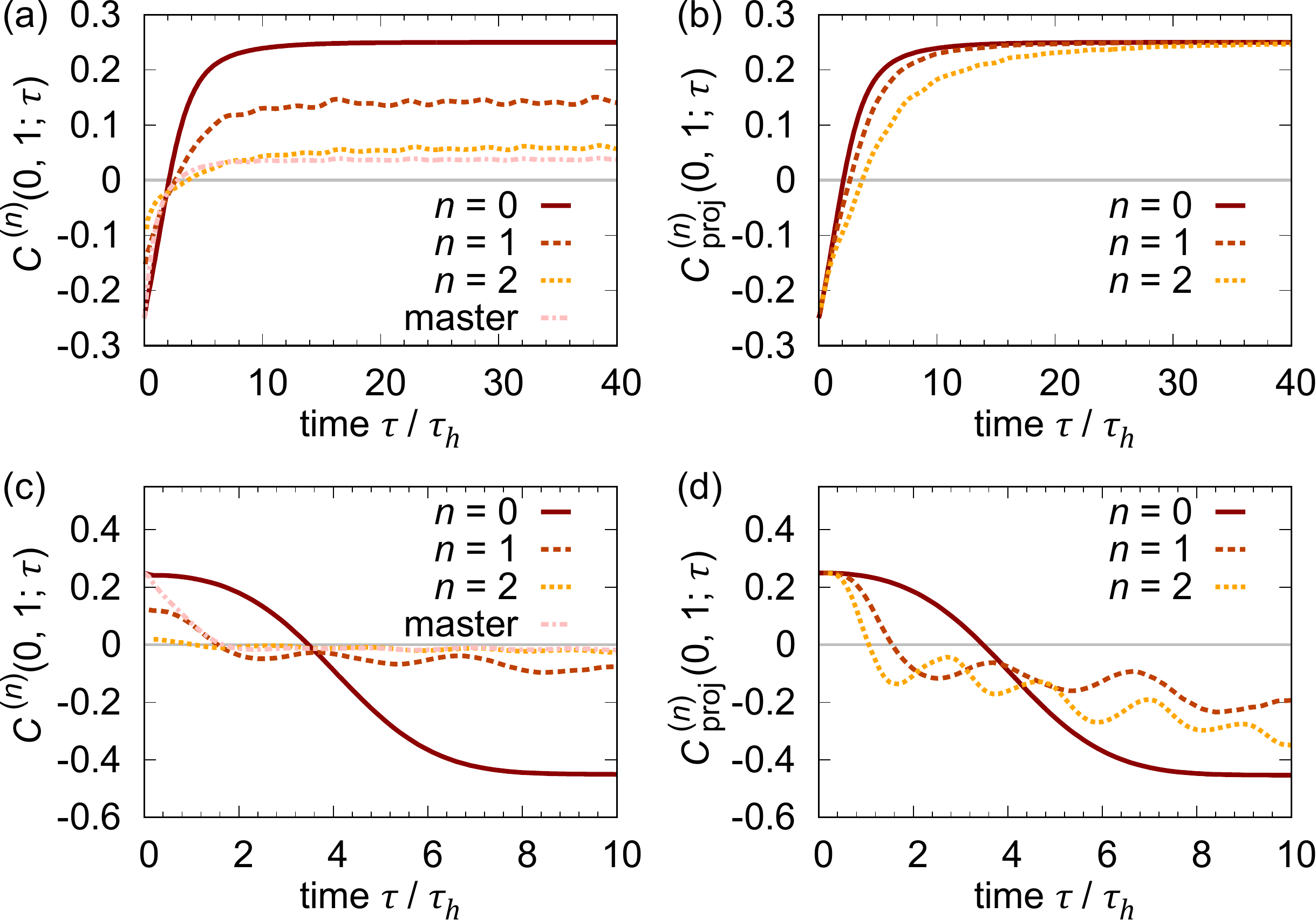}
\caption{(a),(c) Dynamics of spin correlations $C^{(n)}(0,1;\tau)$ averaged over quantum trajectories that involve $n$ quantum jumps. The label ``master'' corresponds to $C(0,1;\tau)$, in which the correlation is calculated from the full density matrix of the solution to the master equation. (b),(d) Dynamics of conditional correlators $C^{(n)}_{\mathrm{proj}}(0,1;\tau)$ after elimination of the effect of holes by additional projection. (a) and (b) The results for the dissipative Fermi-Hubbard model; (c) and (d) those for the dissipative Bose-Hubbard model. The parameters and the initial states are the same as in Fig.~\ref{fig_nojump}. The unit of time is the inverse hopping rate $\tau_h=1/t$.}
\label{fig_withjump}
\end{figure}

Figure \ref{fig_withjump}(a) [\ref{fig_withjump}(c)] shows the dynamics of the magnetic correlation $C^{(n)}(0,1;\tau)$ of the dissipative Fermi (Bose) Hubbard system. 
For comparison, we also show $C(0,1;\tau)\equiv\mathrm{Tr}[\rho(\tau)\bm{S}_0\cdot\bm{S}_1]$, where the average is taken over all quantum trajectories so as to give the solution of the master equation \eqref{eq_master}. The result indicates that the sign reversal of the magnetic correlations is still seen in the presence of quantum jumps, and the magnitude of the correlation increases with decreasing the number of quantum jumps.

%%%%%%%%%%%%%%%%%%%%%%%%%
%%%%%%%%%%%%%%%%%%%%%%%%%

The correlation function $C^{(n)}(i,j;\tau)$ includes the effect of holes produced by quantum jumps. However, one can remove the effect of holes and extract the contribution from spins remaining in the system by imposing a further condition with the following conditional correlator \cite{supple}:
\begin{equation}
C_{\mathrm{proj}}^{(n)}(j,j+1;\tau)\equiv\frac{\mathrm{Tr}[P_jP_{j+1}\rho^{(n)}(\tau)P_jP_{j+1}\bm{S}_j\cdot\bm{S}_{j+1}]}{\mathrm{Tr}[P_jP_{j+1}\rho^{(n)}(\tau)P_jP_{j+1}]},
\end{equation}
where $P_j$ is a projector onto states in which site $j$ is singly occupied. More generally, one can use a correlation function $C^{(n)}_{\mathrm{proj}}(j,j+d;d_h;\tau)\equiv\mathrm{Tr}[P_jQ_{d_h}P_{j+d}\rho^{(n)}(\tau)P_jQ_{d_h}P_{j+d}\bm{S}_j\cdot\bm{S}_{j+d}]/\mathrm{Tr}[P_jQ_{d_h}P_{j+d}\rho^{(n)}(\tau)P_jQ_{d_h}P_{j+d}]$, 
where $Q_{d_h}$ is another projector onto states with $d_h$ holes and $d-d_h-1$ singly occupied sites between sites $j$ and $j+d$. Such conditional correlators have been measured with quantum-gas microscopy \cite{Endres11, Hilker17} by collecting images that match the conditions. 

Numerical results of the conditional correlators $C^{(n)}_{\mathrm{proj}}(0,1;\tau)$ for the Fermi (Bose) Hubbard system are shown in Fig.~\ref{fig_withjump}(b) [\ref{fig_withjump}(d)]. Notably, the magnetic correlations are significantly enhanced from those without projection and even saturated at the same maximum value as in the case without quantum jumps for the Fermi-Hubbard system. While saturation is not achieved in the Bose-Hubbard system since the numerical simulation is limited to $\tau/\tau_h\lesssim 10$ for sufficient statistical convergence, similar saturation behavior can be seen at a single-trajectory level \cite{supple}. Nevertheless, a significant increase in the antiferromagnetic correlation is clearly seen by comparison between Figs.~\ref{fig_withjump}(c) and \ref{fig_withjump}(d). 

The underlying physics behind these results is spin-charge separation in one-dimensional systems \cite{Giamarchi_book}. 
In the strongly correlated Hubbard chain, the created holes move freely as if they were noninteracting, while the background spin state remains the same as that of the Heisenberg chain \cite{Ogata90}. In particular, given an eigenstate of the one-dimensional Hubbard chain, one can reconstruct an eigenstate of the Heisenberg model by eliminating holes involved in individual particle configurations that are superposed in the quantum state \cite{Hilker17, Ogata90, Kruis04}. Thus, the conditional correlators $C^{(n)}_{\mathrm{proj}}(j,j+1;\tau)$ and $C^{(n)}_{\mathrm{proj}}(j,j+d;d_h;\tau)$ capture the essential features of spin correlations in the background Heisenberg model, which are equivalent to those in the case without holes at least in the highest-energy spin state that can be achieved in the long-time limit. 
This explains the saturated value of the conditional spin correlation that exactly coincides with that in the trajectory without loss events shown in Fig.~\ref{fig_nojump}. Although the original argument on the spin-charge separation in eigenstates of the Hubbard model was limited to the fermion case \cite{Hilker17, Ogata90, Kruis04}, our numerical results indicate that this mechanism also works for the Bose-Hubbard system.

%%%%%[Summary]%%%%%

\textit{Summary and future perspectives}.--\ 
We have shown that the inelastic Hubbard interaction alters the spin-exchange process due to a finite lifetime of the intermediate state, leading to novel quantum magnetism opposite to the conventional equilibrium magnetism. Rather than stabilizing low-energy states, high-energy spin states have longer lifetimes and are thus realized in dissipative systems. The Hubbard models with inelastic interactions can be realized with various types of ultracold atoms with internal excited states. A possible experimental platform is a system of ytterbium atoms having long-lived excited states for which the decay to the ground state due to spontaneous emission is negligible \cite{Sponselee18, Tomita18}. Furthermore, inelastic collisions can be artificially induced by using photoassociation techniques \cite{Tomita17}, which will enable the control of quantum magnetism with dissipation.

Our work raises interesting questions for future investigation. First, while we have shown that the effect of holes can be eliminated in one-dimensional systems due to spin-charge separation, it cannot in two (or higher) dimensions. Second, since the Bose-Hubbard system develops antiferromagnetic correlations due to dissipation, geometric frustration in the lattice may realize quantum spin liquids and topological order, which have not yet been realized in cold-atom experiments due to the difficulty of cooling. Third, if the spin SU(2) symmetry is relaxed, eigenstates of the non-Hermitian spin Hamiltonian with the complex-valued spin-exchange couplings are no longer the same as those of the original Hermitian spin Hamiltonian. It is therefore worthwhile to explore novel quantum magnetism in these non-Hermitian spin Hamiltonians \cite{Lee14}.

%\begin{acknowledgments}
%\textit{Acknowledgments}.--\ 
We thank Kazuya Fujimoto, Takeshi Fukuhara, and Yoshiro Takahashi for helpful discussions. 
This work was supported by KAKENHI (Grants No.~JP16K05501, No.~JP16K17729, No.~JP18H01140, No.~JP18H01145, and No.~JP19H01838) and a Grant-in-Aid for Scientific Research on Innovative Areas (KAKENHI Grant No.~JP15H05855) from the Japan Society for the Promotion of Science. 
M.N. was supported by RIKEN Special Postdoctoral Researcher Program. 
N.T. acknowledges support by JST PRESTO (Grant No.~JPMJPR16N7).
\bibliography{NTMag_ref.bib}

%%%%%%%[Supplemental Material]%%%%%%%%%%

\clearpage

\renewcommand{\thesection}{S\arabic{section}}
\renewcommand{\theequation}{S\arabic{equation}}
\setcounter{equation}{0}
\renewcommand{\thefigure}{S\arabic{figure}}
\setcounter{figure}{0}

\onecolumngrid
\appendix
\begin{center}
\large{Supplemental Material for}\\
\textbf{``Dynamical Sign Reversal of Magnetic Correlations in Dissipative Hubbard Models''}
\end{center}
%\maketitle
%\onecolumngrid
%\appendix

%%%-----[Derivation of non-Hermitian spin Hamiltonian]-----%%%

\section{Non-Hermitian spin Hamiltonian}
We derive the non-Hermitian spin Hamiltonian that governs the time evolution in a strongly correlated regime. We start with an effective non-Hermitian Hubbard Hamiltonian $H_{\mathrm{eff}}$ and decompose it into the kinetic part $H'$ and the interaction part $H_0$, where
\begin{align}
H'&=-t\sum_{\langle i,j\rangle}\sum_{\sigma=\uparrow,\downarrow}(c_{i\sigma}^\dag c_{j\sigma}+\mathrm{H.c.}),\notag\\
H_0&=(U-i\gamma)\sum_j n_{j\uparrow}^{(f)}n_{j\downarrow}^{(f)},\notag
\end{align}
for fermions, and
\begin{align}
H'&=-t\sum_{\langle i,j\rangle}\sum_{\sigma=\uparrow,\downarrow}(b_{i\sigma}^\dag b_{j\sigma}+\mathrm{H.c.}),\notag\\
H_0&=(U_{\uparrow\downarrow}-i\gamma_{\uparrow\downarrow})\sum_j n_{j\uparrow}^{(b)}n_{j\downarrow}^{(b)}+\sum_j\sum_{\sigma=\uparrow,\downarrow}\frac{U_{\sigma\sigma}-i\gamma_{\sigma\sigma}}{2}n_{j\sigma}^{(b)}(n_{j\sigma}^{(b)}-1),\notag
\end{align}
for bosons. In the strongly correlated regime $U,U_{\sigma\sigma'}\gg t$, the kinetic term $H'$ can be treated as a perturbation. For simplicity, we consider a Mott insulating state and ignore holes. 
According to the second-order perturbation theory, an effective Hamiltonian is given by
\begin{equation}
H_{\mathrm{spin}}=E_0+\mathcal{P}H'\frac{1}{E_0-H_0}H'\mathcal{P},
\end{equation}
where $\mathcal{P}$ is a projector onto the Hilbert subspace in which each lattice site is occupied by one atom. Here the energy $E_0$ of the unperturbed state is set to $E_0=0$. In the simplest two-site case, the Hilbert subspace is spanned by four spin configurations $\{\ket{\uparrow\uparrow},\ket{\uparrow\downarrow},\ket{\downarrow\uparrow},\ket{\downarrow\downarrow}\}$. In this case, the spin Hamiltonian reads
\begin{align}
H_{\mathrm{spin}}=-\frac{2t^2}{U-i\gamma}(\ket{\uparrow\downarrow}\bra{\uparrow\downarrow}+\ket{\downarrow\uparrow}\bra{\downarrow\uparrow}-\ket{\downarrow\uparrow}\bra{\uparrow\downarrow}-\ket{\uparrow\downarrow}\bra{\downarrow\uparrow}),
\end{align}
for fermions, and
\begin{align}
H_{\mathrm{spin}}=-2t^2\Bigl(&\frac{2}{U_{\uparrow\uparrow}-i\gamma_{\uparrow\uparrow}}\ket{\uparrow\uparrow}\bra{\uparrow\uparrow}+\frac{2}{U_{\downarrow\downarrow}-i\gamma_{\downarrow\downarrow}}\ket{\downarrow\downarrow}\bra{\downarrow\downarrow}+\frac{1}{U_{\uparrow\downarrow}-i\gamma_{\uparrow\downarrow}}\ket{\uparrow\downarrow}\bra{\uparrow\downarrow}+\frac{1}{U_{\uparrow\downarrow}-i\gamma_{\uparrow\downarrow}}\ket{\downarrow\uparrow}\bra{\downarrow\uparrow}\notag\\
&+\frac{1}{U_{\uparrow\downarrow}-i\gamma_{\uparrow\downarrow}}\ket{\downarrow\uparrow}\bra{\uparrow\downarrow}+\frac{1}{U_{\uparrow\downarrow}-i\gamma_{\uparrow\downarrow}}\ket{\uparrow\downarrow}\bra{\downarrow\uparrow}\Bigr),
\end{align}
for bosons. Hence, for fermions, the spin Hamiltonian is given by the non-Hermitian Heisenberg model
\begin{equation}
H_{\mathrm{spin}}=\frac{4t^2}{U-i\gamma}\sum_{\langle i,j\rangle}\left(\bm{S}_i\cdot\bm{S}_j-\frac{1}{4}\right),
\label{eq_Hspin_fermion}
\end{equation}
and for bosons it is given by
\begin{align}
H_{\mathrm{spin}}=&\sum_{\langle i,j\rangle}\Bigl[-\frac{4t^2}{U_{\uparrow\downarrow}-i\gamma_{\uparrow\downarrow}}(S_i^xS_j^x+S_i^yS_j^y)-4t^2\Bigl(\frac{1}{U_{\uparrow\uparrow}-i\gamma_{\uparrow\uparrow}}+\frac{1}{U_{\downarrow\downarrow}-i\gamma_{\downarrow\downarrow}}-\frac{1}{U_{\uparrow\downarrow}-i\gamma_{\uparrow\downarrow}}\Bigr)S_i^zS_j^z\notag\\
&-t^2\Bigl(\frac{1}{U_{\uparrow\uparrow}-i\gamma_{\uparrow\uparrow}}+\frac{1}{U_{\downarrow\downarrow}-i\gamma_{\downarrow\downarrow}}+\frac{1}{U_{\uparrow\downarrow}-i\gamma_{\uparrow\downarrow}}\Bigr)-2t^2\Bigl(\frac{1}{U_{\uparrow\uparrow}-i\gamma_{\uparrow\uparrow}}-\frac{1}{U_{\downarrow\downarrow}-i\gamma_{\downarrow\downarrow}}\Bigr)(S_i^z+S_j^z)\Bigr]\notag\\
=&\sum_{\langle i,j\rangle}\Bigl[(J_{\mathrm{eff}}^\perp+i\Gamma^\perp)(S_i^xS_j^x+S_i^yS_j^y)+(J_{\mathrm{eff}}^z+i\Gamma^z)S_i^zS_j^z+C\Bigr]+(h_r+ih_i)\sum_j S_j^z,
\label{eq_Hspin_boson}
\end{align}
where
\begin{align}
J_{\mathrm{eff}}^\perp&=-\mathrm{Re}\left[\frac{4t^2}{U_{\uparrow\downarrow}-i\gamma_{\uparrow\downarrow}}\right],\\
\Gamma^\perp&=-\mathrm{Im}\left[\frac{4t^2}{U_{\uparrow\downarrow}-i\gamma_{\uparrow\downarrow}}\right],\\
J_{\mathrm{eff}}^z&=-4t^2\mathrm{Re}\left[\frac{1}{U_{\uparrow\uparrow}-i\gamma_{\uparrow\uparrow}}+\frac{1}{U_{\downarrow\downarrow}-i\gamma_{\downarrow\downarrow}}-\frac{1}{U_{\uparrow\downarrow}-i\gamma_{\uparrow\downarrow}}\right],\\
\Gamma^z&=-4t^2\mathrm{Im}\left[\frac{1}{U_{\uparrow\uparrow}-i\gamma_{\uparrow\uparrow}}+\frac{1}{U_{\downarrow\downarrow}-i\gamma_{\downarrow\downarrow}}-\frac{1}{U_{\uparrow\downarrow}-i\gamma_{\uparrow\downarrow}}\right],\\
C&=-t^2\Bigl(\frac{1}{U_{\uparrow\uparrow}-i\gamma_{\uparrow\uparrow}}+\frac{1}{U_{\downarrow\downarrow}-i\gamma_{\downarrow\downarrow}}+\frac{1}{U_{\uparrow\downarrow}-i\gamma_{\uparrow\downarrow}}\Bigr),\\
h_r&=-2z_ct^2\mathrm{Re}\left[\frac{1}{U_{\uparrow\uparrow}-i\gamma_{\uparrow\uparrow}}-\frac{1}{U_{\downarrow\downarrow}-i\gamma_{\downarrow\downarrow}}\right],\\
h_i&=-2z_ct^2\mathrm{Im}\left[\frac{1}{U_{\uparrow\uparrow}-i\gamma_{\uparrow\uparrow}}-\frac{1}{U_{\downarrow\downarrow}-i\gamma_{\downarrow\downarrow}}\right].
\end{align}
Here, $z_c$ denotes the coordination number of the lattice. For $U_{\uparrow\uparrow}=U_{\downarrow\downarrow}=U_{\uparrow\downarrow}=U$ and $\gamma_{\uparrow\uparrow}=\gamma_{\downarrow\downarrow}=\gamma_{\uparrow\downarrow}=\gamma$, the model for bosons \eqref{eq_Hspin_boson} reduces to the non-Hermitian Heisenberg model considered in the main text. If a bosonic system does not respect the spin SU($2$) symmetry, the non-Hermitian spin model \eqref{eq_Hspin_boson} is an XXZ model with complex-valued interaction strength and a magnetic field. We note that the effective magnetic field has an imaginary part $h_i$ in general. In the Hermitian case, the real magnetic field $h_r$ can be compensated by an additional external magnetic field \cite{Duan03}. However, the imaginary magnetic field cannot be compensated by any real external field and thus inevitably affects the behavior of dissipative spin systems.

%%%-----[Dependnce on dissipation]-----%%%

\section{Dependence of the dynamics on dissipation}
\begin{figure}
\includegraphics[width=10cm]{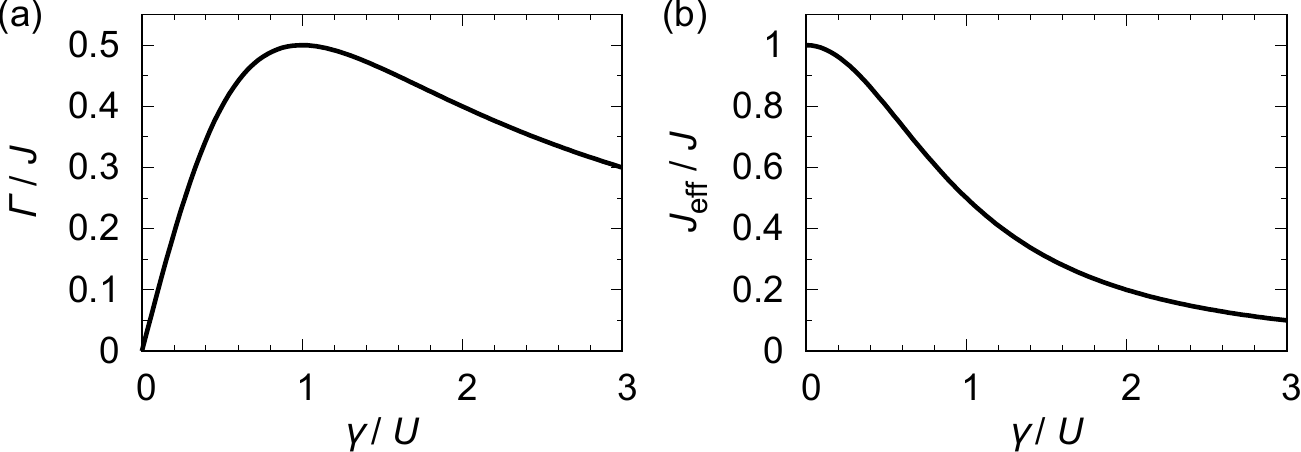}
\caption{(a) $\gamma$-dependence of the imaginary part of the effective spin-exchange interaction. (b) $\gamma$-dependence of the real part of the effective spin-exchange interaction. Here $J$ is given by $J=4t^2/U$.}
\label{fig_Jeff}
\end{figure}

In Fig.~\ref{fig_Jeff}, we show how the real and imaginary parts of the effective spin-exchange interactions, which are respectively given by $J_{\mathrm{eff}}=4Ut^2/(U^2+\gamma^2)$ and $\Gamma=4\gamma t^2/(U^2+\gamma^2)$, depend on the inelastic collision rate $\gamma$. The imaginary part reaches the maximum $\Gamma=0.5J$ at $\gamma/U=1$ and then decreases with increasing $\gamma$. The suppression of the effective dissipation rate $\Gamma$ at large $\gamma$ is attributed to the continuous quantum Zeno effect \cite{Syassen08, Zhu14, Tomita17}, which freezes the hopping of atoms due to a large dissipation. On the other hand, the real part $J_{\mathrm{eff}}$ of the spin-exchange interaction monotonically decreases as a function of $\gamma$.

\begin{figure}
\includegraphics[width=17.5cm]{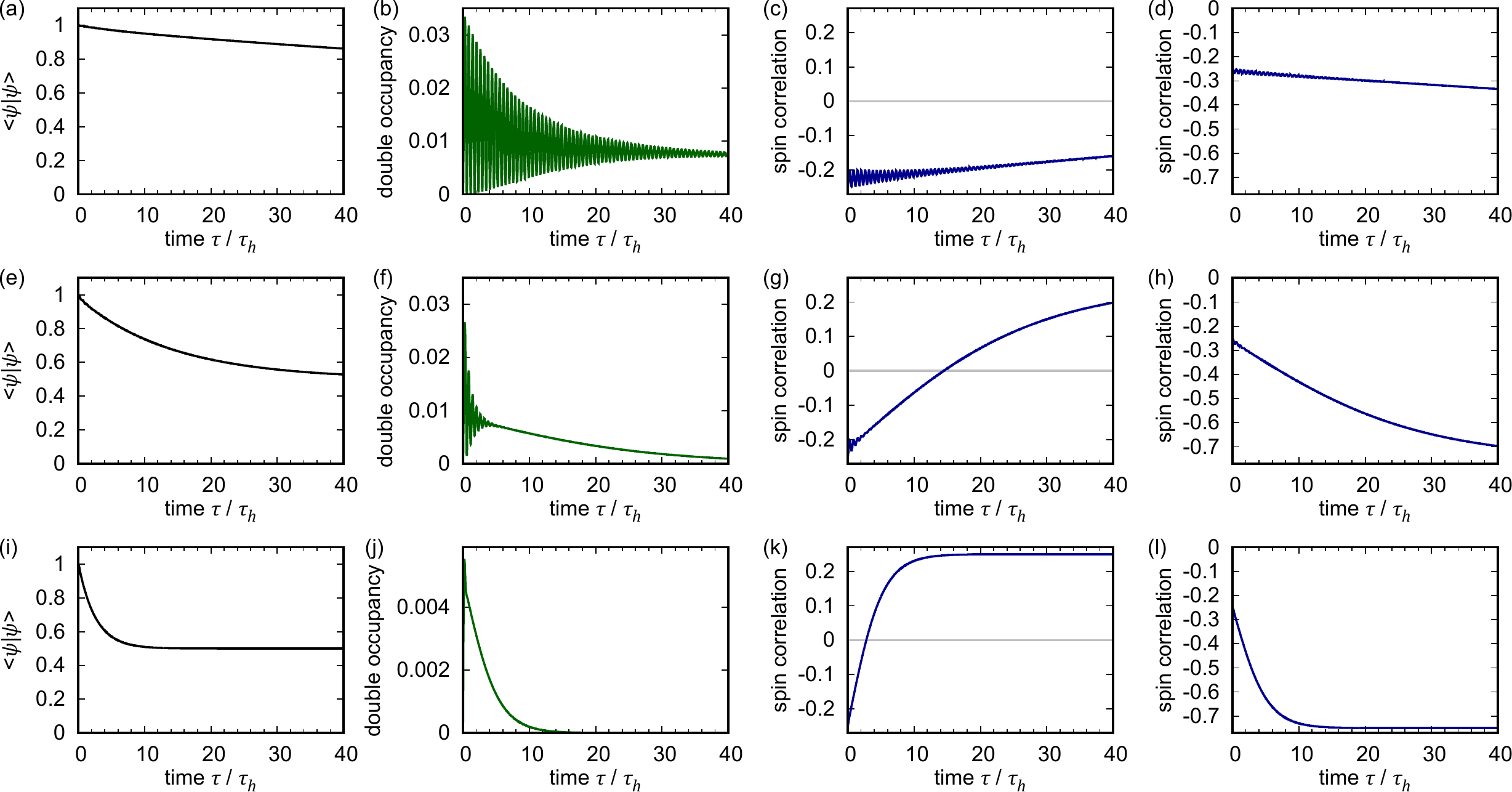}
\caption{Dependence of the dynamics of the two-site dissipative Hubbard model on dissipation. (a), (e), (i) Time evolution of the squared norm $\braket{\psi(\tau)|\psi(\tau)}$. (b), (f), (j) Time evolution of the double occupancy $\bra{\psi(\tau)}\frac{1}{2}(n_{1\up}^{(a)}n_{1\down}^{(a)}+n_{2\up}^{(a)}n_{2\down}^{(a)})\ket{\psi(\tau)}/\braket{\psi(\tau)|\psi(\tau)}$ ($a=f$ or $b$). The squared norm and the double occupancy take the same values for the Fermi and Bose Hubbard systems. (c), (g), (k) Time evolution of the spin correlation $\bra{\psi(\tau)}\bm{S}_1\cdot\bm{S}_2\ket{\psi(\tau)}/\braket{\psi(\tau)|\psi(\tau)}$ of the non-Hermitian Fermi-Hubbard model. (d), (h), (l) Time evolution of the spin correlation of the non-Hermitian Bose-Hubbard model. The strength of the interaction is set to $U/t=10$ in all figures, and the strength of dissipation is set to $\gamma/t=0.1$ in (a)-(d), $\gamma/t=1$ in (e)-(h), and $\gamma/t=10$ in (i)-(l). The unit of time is the inverse hopping rate $\tau_h=1/t$.}
\label{fig_2site_supple}
\end{figure}

The dependence of the dynamics of the Hubbard model on dissipation is shown in Fig.~\ref{fig_2site_supple}. Here we calculate the dynamics of the two-site non-Hermitian Fermi and Bose Hubbard models which can be realized with a double-well optical lattice as mentioned in the main text. When small dissipation is introduced to the system [Figs.~\ref{fig_2site_supple}(a)-(d)], fast oscillations of the double occupancy and the spin correlation due to a large on-site repulsion $U$ are damped by dissipation. As the strength of dissipation is increased [Figs.~\ref{fig_2site_supple}(e)-(h)], the development of the ferromagnetic (antiferromagnetic) spin correlation in the Fermi (Bose) system is accelerated by an increase in the imaginary part of the spin-exchange interaction $\Gamma$, which governs the time scale of the dissipative spin dynamics. At the optimal value $\gamma/U=1$ [Figs.~\ref{fig_2site_supple}(i)-(l)], the fastest formation of the spin correlation is observed. We note that the double occupancy is gradually suppressed with increasing the dissipation [see Figs.~\ref{fig_2site_supple}(b), (f), and (j)]. This behavior is a consequence of the continuous quantum Zeno effect, as mentioned in the main text.

%%%%%%%%%%%%%%%%%%%%%%%%%
%%%%%%%%%%%%%%%%%%%%%%%%%

%%%%%%%%%%%%%%%%%%%%%%%%%
%%%%%%%%%%%%%%%%%%%%%%%%%

%%%-----[Quantum-trajectory method]-----%%%

\section{Details of the quantum-trajectory method}
The dynamics of a dissipative Hubbard model is simulated by using a quantum-trajectory method \cite{Dalibard92, Carmichael_book, Daley14}. According to a random number $R_1$ chosen from an interval $0\leq R_1\leq 1$, the system evolves under the nonunitary Schr\"{o}dinger equation $i\partial_\tau\ket{\tilde{\psi}(\tau)}=H_{\mathrm{eff}}\ket{\tilde{\psi}(\tau)}$ up to a time $\tau_1$ when the squared norm $\braket{\tilde{\psi}(\tau_1)|\tilde{\psi}(\tau_1)}$ is equal to $R_1$. Here $H_{\mathrm{eff}}$ is the $N$-site non-Hermitian Fermi or Bose Hubbard Hamiltonian and we assume the periodic boundary condition. The Schr\"{o}dinger dynamics is numerically calculated by exact diagonalization of $H_{\mathrm{eff}}$. At time $\tau_1$, a loss event takes place and the state is acted on by the quantum-jump operator $L_{j\sigma\sigma'}$ and then normalized: 
\begin{equation}
\ket{\tilde{\psi}(\tau_1+0)}=\frac{L_{j\sigma\sigma'}\ket{\tilde{\psi}(\tau_1-0)}}{\sqrt{\bra{\tilde{\psi}(\tau_1-0)}L_{j\sigma\sigma'}^\dag L_{j\sigma\sigma'}\ket{\tilde{\psi}(\tau_1-0)}}}.
\end{equation}
The quantum-jump operator $L_{j\sigma\sigma'}$ for the loss event at $\tau_1$ is chosen according to the probability distribution
\begin{equation}
\frac{\bra{\tilde{\psi}(\tau_1-0)}L_{j\sigma\sigma'}^\dag L_{j\sigma\sigma'}\ket{\tilde{\psi}(\tau_1-0)}}{\sum_{j,\sigma,\sigma'}\bra{\tilde{\psi}(\tau_1-0)}L_{j\sigma\sigma'}^\dag L_{j\sigma\sigma'}\ket{\tilde{\psi}(\tau_1-0)}}.
\end{equation}
After $\tau_1$, we take another random number $R_2$ and repeat the above procedure. 
When a sufficiently large number $\mathcal{N}$ of quantum trajectories are sampled, the density matrix of the solution of the master equation (3) is given by
\begin{equation}
\rho(\tau)\simeq \frac{1}{\mathcal{N}}\sum_{a=1}^{\mathcal{N}}\ket{\psi_a(\tau)}\bra{\psi_a(\tau)},
\label{eq_trajrho}
\end{equation}
where $\ket{\psi_a(\tau)}=\ket{\tilde{\psi}_a(\tau)}/\sqrt{\braket{\tilde{\psi}_a(\tau)|\tilde{\psi}_a(\tau)}}$ is a normalized state along the $a$-th quantum trajectory $(a=1,\cdots,\mathcal{N})$. The approximate equality becomes the exact one in the limit of $\mathcal{N}\to\infty$.

Each quantum trajectory can be characterized by the number of quantum jumps. Let $N^{(n)}(\tau)$ be the number of quantum trajectories that involve $n$ quantum jumps between the initial time and $\tau$. Then, from Eq.~\eqref{eq_trajrho}, the density matrix conditioned on the number of quantum jumps is given by
\begin{align}
\rho^{(n)}(\tau)=&\frac{\mathcal{P}^{(n)}\rho(\tau)\mathcal{P}^{(n)}}{\mathrm{Tr}[\mathcal{P}^{(n)}\rho(\tau)\mathcal{P}^{(n)}]}\notag\\
\simeq&\frac{1}{N^{(n)}(\tau)}\sum_{a=1}^{N^{(n)}(\tau)}\ket{\psi^{(n)}_a(\tau)}\bra{\psi^{(n)}_a(\tau)},
\end{align}
where $\ket{\psi^{(n)}_a(\tau)}\ (a=1,\cdots,N^{(n)}(\tau))$ denotes the normalized state along the $a$-th quantum trajectory that includes $n$ quantum jumps. The correlation function is thus calculated as
\begin{align}
C^{(n)}(i,j;\tau)=&\mathrm{Tr}[\rho^{(n)}(\tau)\bm{S}_i\cdot\bm{S}_j]\notag\\
\simeq&\frac{1}{N^{(n)}(\tau)}\sum_{a=1}^{N^{(n)}(\tau)}\bra{\psi^{(n)}_a(\tau)}\bm{S}_i\cdot\bm{S}_j\ket{\psi^{(n)}_a(\tau)}.
\end{align}
Similarly, the conditional correlator $C^{(n)}_{\mathrm{proj}}(j,j+1;\tau)$ can also be calculated as
\begin{align}
C_{\mathrm{proj}}^{(n)}(j,j+1;\tau)=&\frac{\mathrm{Tr}[P_jP_{j+1}\rho^{(n)}(\tau)P_jP_{j+1}\bm{S}_j\cdot\bm{S}_{j+1}]}{\mathrm{Tr}[P_jP_{j+1}\rho^{(n)}(\tau)P_jP_{j+1}]}\notag\\
\simeq&\frac{\sum_{a=1}^{N^{(n)}(\tau)}\bra{\psi^{(n)}_a(\tau)}P_jP_{j+1}\bm{S}_j\cdot\bm{S}_{j+1}P_jP_{j+1}\ket{\psi^{(n)}_a(\tau)}}{\sum_{a=1}^{N^{(n)}(\tau)}\bra{\psi^{(n)}_a(\tau)}P_jP_{j+1}\ket{\psi^{(n)}_a(\tau)}}.
\end{align}

\begin{figure}
\includegraphics[width=12cm]{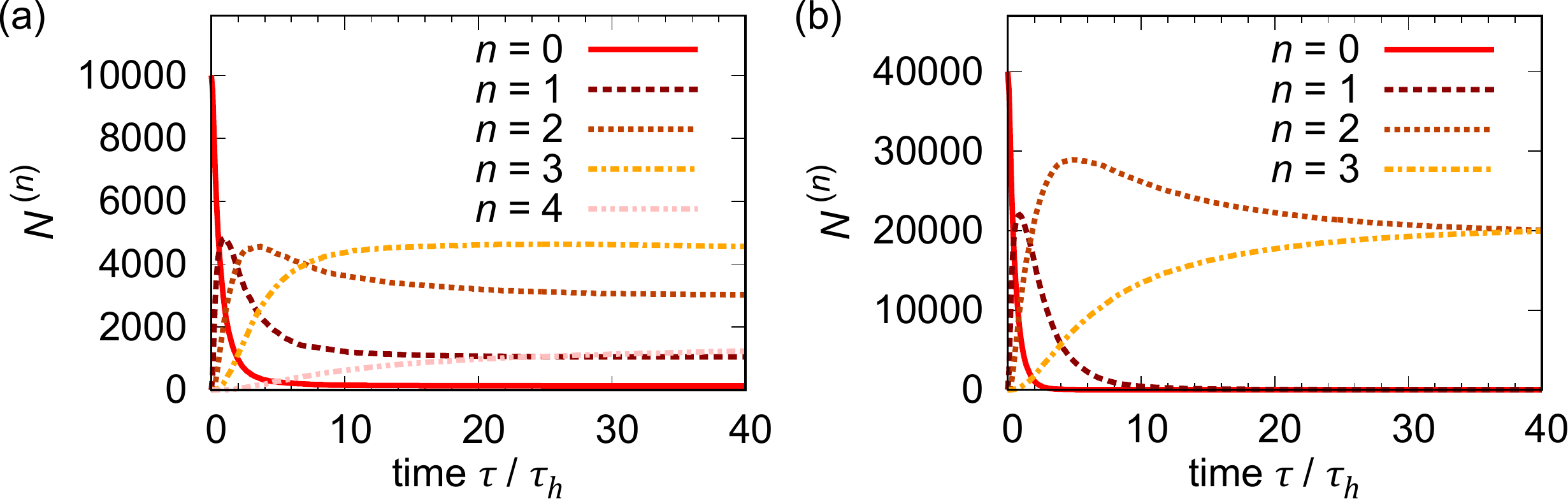}
\caption{Time evolution of the number of quantum trajectories $N^{(n)}(\tau)$ that include $n$ quantum jumps between the initial time and $\tau$ for (a) the $8$-site dissipative Fermi-Hubbard model and (b) the $6$-site dissipative Bose-Hubbard model. The parameters and the initial states are the same as in Fig.~\ref{fig_withjump} in the main text.}
\label{fig_Ntraj}
\end{figure}

In the numerical simulation, we use $\mathcal{N}=10000$ trajectories for the $8$-site dissipative Fermi-Hubbard model and $\mathcal{N}=40000$ trajectories for the $6$-site dissipative Bose-Hubbard model. In Fig.~\ref{fig_Ntraj}, we show the time evolution of the number of quantum trajectories $\mathcal{N}^{(n)}(\tau)$. For the case of the Fermi-Hubbard system, $N^{(n)}(\tau)$ for each $n$ remains nonvanishing even after a long time since the Fermi-Hubbard system has dark states, which are spin-symmetric Dicke states \cite{FossFeig12}, in each particle-number sector. In particular, we have $N^{(n=0)}(\tau)\simeq 100$ trajectories with no quantum jump at $\tau/\tau_h=40$. On the other hand, the Bose-Hubbard system does not have a dark state except for the two-particle sector which corresponds to the $n=2$ case in Fig.~\ref{fig_Ntraj}(b), since $N$ spins cannot form a perfect antisymmetric state except for $N=2$. As a result, $N^{(n=0)}(\tau)$ and $N^{(n=1)}(\tau)$ in Fig.~\ref{fig_Ntraj}(b) decay and vanish in the long-time limit. To achieve sufficient statistical convergence, we restrict the time to $\tau/\tau_h\lesssim 10$, for which we have $N^{(n=1)}(10\tau_h)\simeq 400$ trajectories.

%%%%%%%%%%%%%%%%%%%%%%%%%
%%%%%%%%%%%%%%%%%%%%%%%%%

%%%-----[Single-trajectory dynamics]-----%%%

\section{Dynamics in single quantum trajectories}
\begin{figure}
\includegraphics[width=12cm]{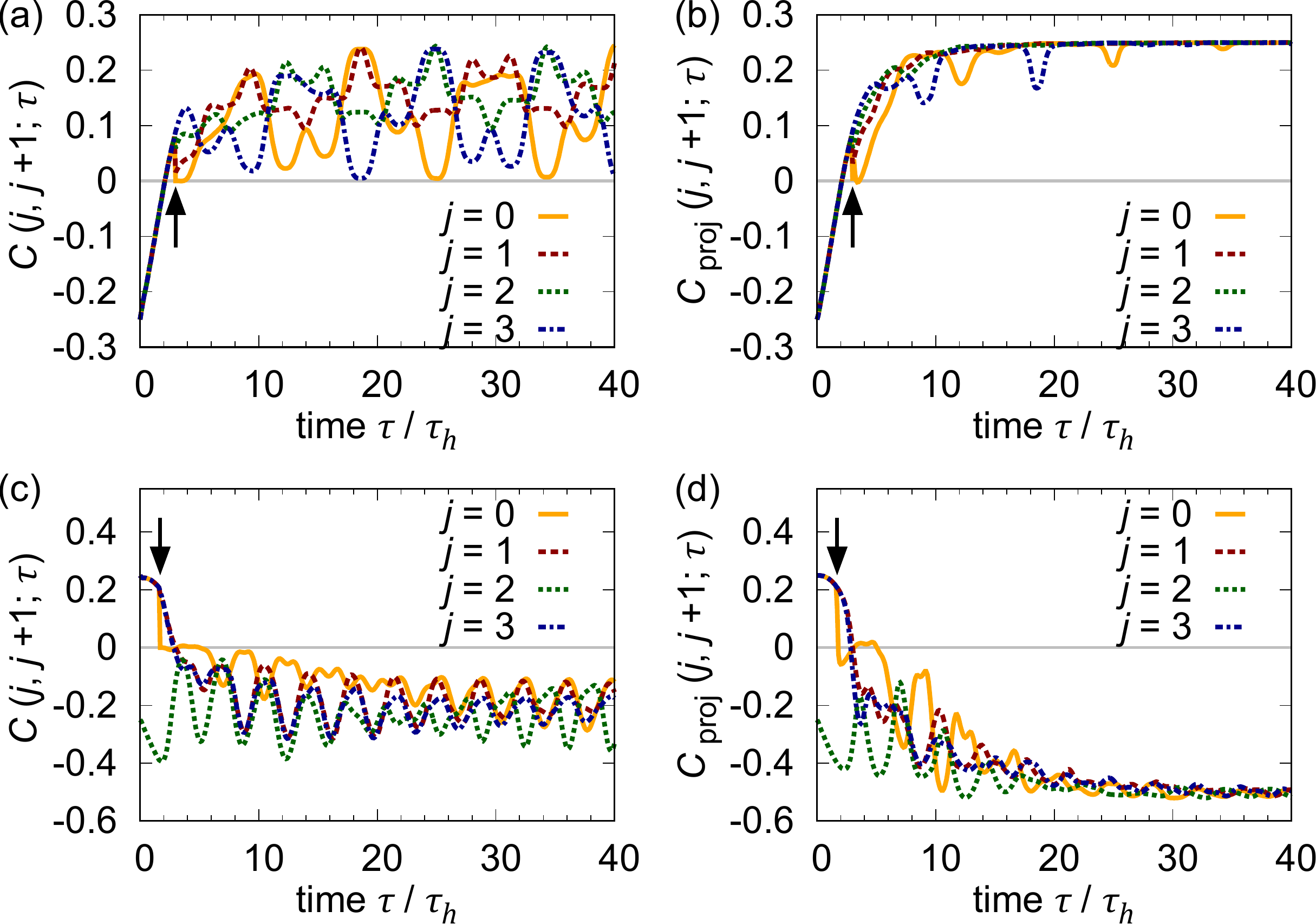}
\caption{(a) (c) Dynamics of spin correlations $C(j,j+1;\tau)$ in a single quantum trajectory which involves a quantum-jump event. (b) (d) Dynamics of conditional spin correlations $C_{\mathrm{proj}}(j,j+1;\tau)$ in the same trajectories. (a) and (b) show the results for the dissipative Fermi-Hubbard model, and (c) and (d) show the results for the dissipative Bose-Hubbard model. The parameters are the same as in Fig.~\ref{fig_nojump}. The arrows indicate the time at which the quantum-jump event takes place. The unit of time is the inverse hopping rate $\tau_h=1/t$.}
\label{fig_traj}
\end{figure}

Figure~\ref{fig_traj}(a) (\ref{fig_traj}(c)) shows the dynamics of the spin correlation $C(j,j+1;\tau)=\bra{\tilde{\psi}(\tau)}\bm{S}_j\cdot\bm{S}_{j+1}\ket{\tilde{\psi}(\tau)}/\braket{\tilde{\psi}(\tau)|\tilde{\psi}(\tau)}$ of the dissipative Fermi (Bose) Hubbard model calculated from a single quantum trajectory which involves a loss event. 
The parameters and the initial states are the same as in Fig.~\ref{fig_nojump}. 
In Fig.~\ref{fig_traj}(a), a quantum-jump event takes place at $\tau/\tau_h\simeq 3$ and creates a hole at site $j=0$. In Fig.~\ref{fig_traj}(c), a quantum-jump event at $\tau/\tau_h\simeq 1.7$ annihilates one spin-up boson and one spin-down boson at site $j=0$. 
In both cases, the spin correlations after the quantum jump oscillate since the created holes move among the lattice sites and disturb the background spin configuration. After the ensemble average is taken, the oscillations disappear, and the spin correlation of the Fermi (Bose) system shows the formation of ferromagnetic (anfiferromagnetic) correlations, while the magnitude is reduced due to the effect of holes [see Figs.~\ref{fig_withjump}(a) and \ref{fig_withjump}(c) in the main text].

In contrast, Fig.~\ref{fig_traj}(b) (\ref{fig_traj}(d)) shows the conditional correlator
\begin{equation}
C_{\mathrm{proj}}(j,j+1;\tau)=\frac{\bra{\psi(\tau)}P_jP_{j+1}\bm{S}_j\cdot\bm{S}_{j+1}P_jP_{j+1}\ket{\psi(\tau)}}{\bra{\psi(\tau)}P_jP_{j+1}\ket{\psi(\tau)}},
\end{equation}
which is calculated from the same trajectories as those in Fig.~\ref{fig_traj}(a) (\ref{fig_traj}(c)).  
Remarkably, although the ferromagnetic (antiferromagnetic) correlation, which develops through the dissipative spin-exchange mechanism, is disturbed by a quantum jump, it starts to grow again and is finally saturated at the same value as in the case of no quantum jump. 
This indicates that the spin configuration after removing holes in the long-time limit is equivalent to that of the highest-energy state of the Heisenberg model as a consequence of spin-charge separation.

%%%%%%%%%%%%%%%%%%%%%%%%%
%%%%%%%%%%%%%%%%%%%%%%%%%

\begin{figure}
\includegraphics[width=12cm]{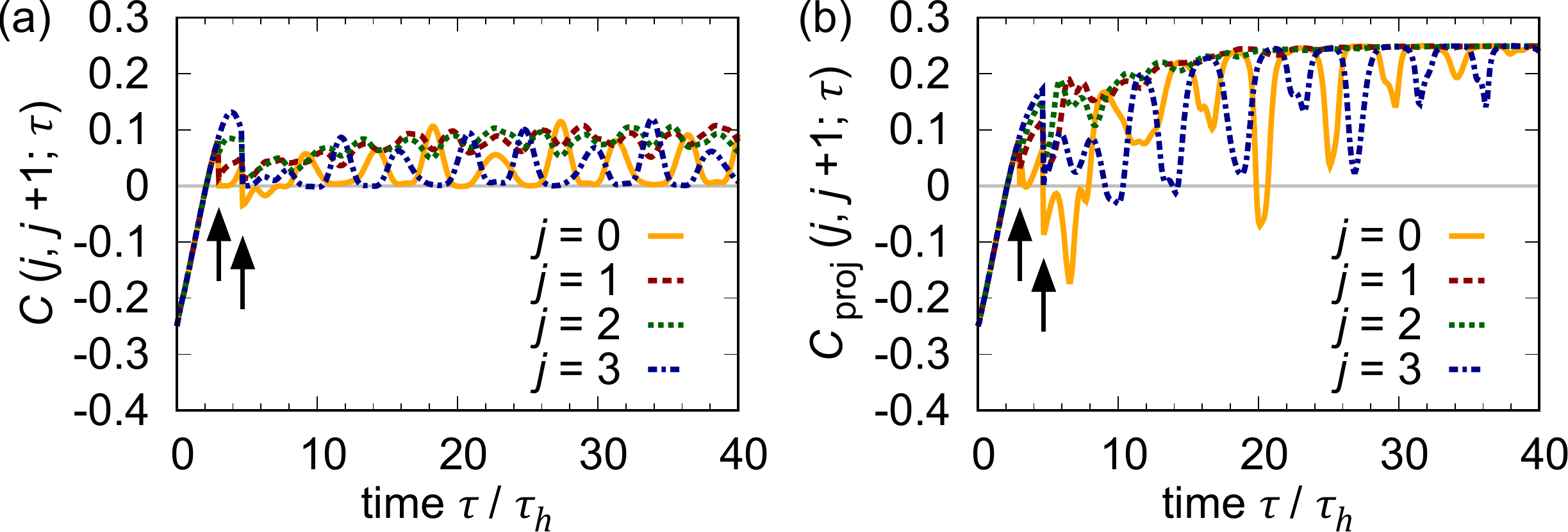}
\caption{(a) Dynamics of spin correlations $C(j,j+1;\tau)$ of the dissipative Fermi-Hubbard model in a quantum trajectory involving two quantum jumps. (b) Dynamics of conditional spin correlations $C_{\mathrm{proj}}(j,j+1;\tau)$ along the same trajectory as that in (a). The parameters are the same as in Fig.~\ref{fig_nojump}. The arrows indicate the times at which the quantum-jump events occur. The unit of time is the inverse hopping rate $\tau_h=1/t$.}
\label{fig_jump2}
\end{figure}

Figure \ref{fig_jump2} shows the dynamics of the spin correlation functions along a quantum trajectory with two jump events. Here the dissipative Fermi-Hubbard model with 8 sites is studied. The initial state is chosen to be the N\'{e}el state as in the main text. The first quantum-jump event at $\tau/\tau_h\simeq 3$ occurs at site $j=0$ and decreases the particle number from eight to six. Subsequently, the second two-body loss event takes place at site $j=4$ at time $\tau/\tau_h\simeq 4.7$, leaving four atoms in the system. As shown in Fig.~\ref{fig_jump2}(b), the conditional correlators involving sites at which the loss events take place are significantly affected by the quantum jumps (see $j=0$ and $j=3$ lines). Remarkably, the conditional correlators at the other sites are not quite disturbed (see $j=1$ and $j=2$ lines) and eventually saturated at the completely ferromagnetic value $C_{\mathrm{proj}}(j,j+1;\tau)=0.25$ in a time scale comparable with that along the quantum trajectory without loss events shown in Fig.~\ref{fig_nojump}(a) in the main text. Such a feature is not clearly observed in the standard correlators [Fig.~\ref{fig_jump2}(a)] and can be probed by the conditional correlators through quantum-gas microscopy.

%%%%%%%%%%%%%%%%%%%%%%%%%
%%%%%%%%%%%%%%%%%%%%%%%%%

\end{document}